# NDE 4.0 – From Design Thinking to Strategy

Johannes VRANA [1] and Ripudaman Singh [2]
[1] Vrana GmbH, Rimsting, Germany
[2] Inspiring Next, Cromwell, CT, USA

Contact E-Mail: contact@vrana.net

## *Abstract*

Cyber technologies are offering new horizons for quality control in manufacturing and safety assurance in-service of physical assets. The line between non-destructive evaluation (NDE) and Industry 4.0 is getting blurred since both are sensory data-driven domains. This multidisciplinary approach has led to the emergence of a new capability: NDE 4.0. The NDT community is coming together once again to define the purpose, chart the process, and address the adoption of emerging technologies.

In this paper, the authors have taken a design thinking approach to spotlight proper objectives for research on this subject. It begins with qualitative research on twenty different perceptions of stakeholders and misconceptions around the current state of NDE. The interpretation is used to define ten value propositions or use cases under 'NDE for Industry 4.0' and 'Industry 4.0 for NDE' leading up to the clarity of purpose for NDE 4.0 – enhanced safety and economic value for stakeholders. To pursue this worthy cause, the paper delves into some of the top adoption challenges, and proposes a journey of managed innovation, conscious skills development, and a new form of leadership required to succeed in the cyber-physical world.

**Notes**: This paper presents the early stages on the road to NDE 4.0. It can also be seen as an orientation to NDE 4.0 to bring awareness and familiarity with the subject, starting with a quick overview of the relevant digital technologies.

## *Introduction*

Historians split recent times into three industrial revolutions: mechanization (steam power), technical (electric power and mass production), and digital (computing and microelectronics). The world of NDE has seen a parallel: first - tools to sharpen human senses, second - wave application to view inside the components, and third - digital processing and automation.

As the industry goes through the fourth revolution powered by interconnections and enhanced digitalization, NDE is also on a new horizon - the NDE 4.0 with the addition of information transparency, technical assistance, machine intelligence, decentralized decisions, and much more.

The term Industrie 4.0 (German for Industry 4.0) was introduced in 2011 at the Hannovermesse in Germany [1] to give a name to all the ongoing activities which will eventually lead to the fourth industrial revolution [2,3,4].

Like the other industrial revolutions (ref to Figure 1), the fourth revolution is shaped by new markets / industries and in particular by new technologies, like digital twin, industrial internet of things, cloud, 5G, augmented reality, artificial intelligence, big data, additive manufacturing, digitalization, robotics and drones, blockchains, and quantum computers. Looking at these emerging technologies connected with Industry 4.0, it becomes clear that this fourth revolution is driven by data and connectivity across the digital and physical worlds.

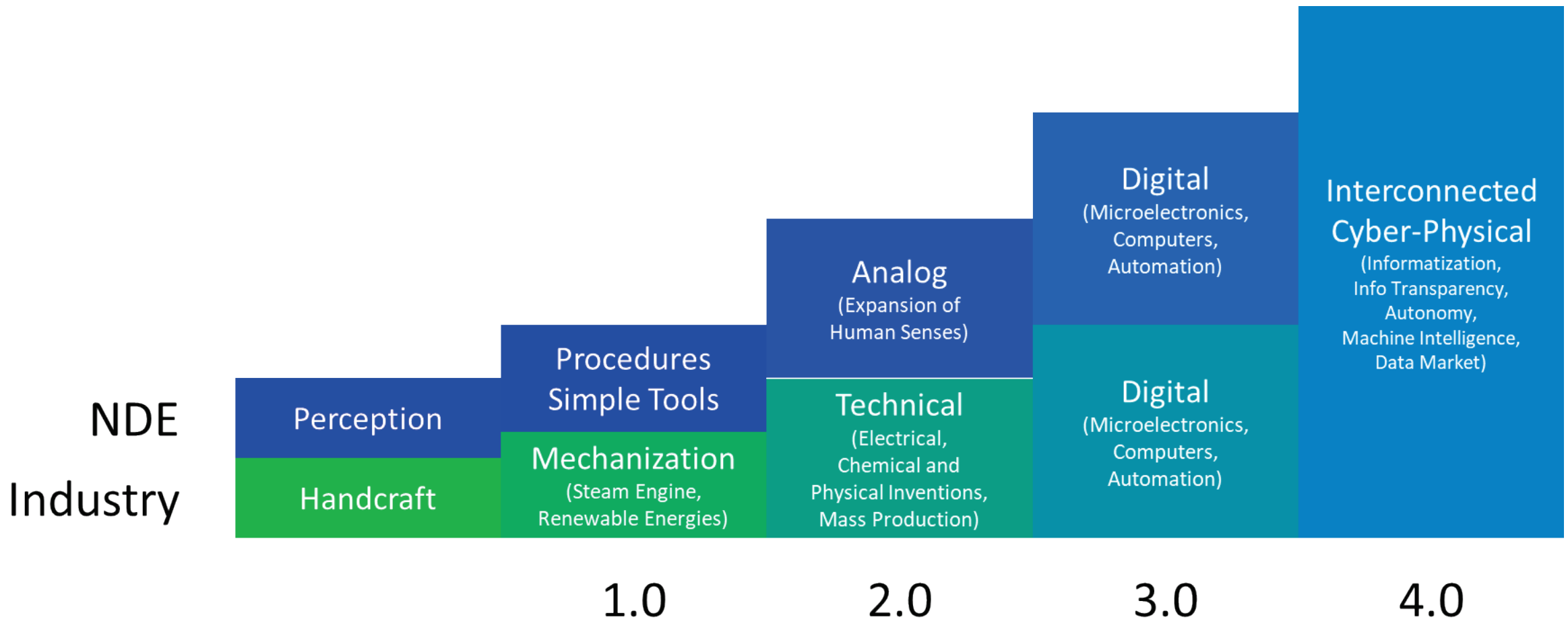

**Figure 1** Visualization of the four industrial and NDE revolutions based on [2,3]. For the definition of the four NDE revolutions it was chosen to define them independent from Industry by the revolutionary changes within NDE. This results in the fact that from strict timetable view the industrial revolutions and the revolutions in NDE did not occur at the same time (in most cases the NDE revolutions are belated) and that the x-Axis of the Figure should not be seen as a strict timeline.
(Author: Johannes Vrana, Vrana GmbH, Licenses: CC BY-ND 4.0)

### *Design Thinking Approach to NDE 4.0*

The amount and pace of technology change, the level of application complexity, and the role of human beings in the NDE system, begs for a dependable approach to development and adoption of NDE 4.0. It needs design thinking - an iterative process which seeks to understand the user, challenge assumptions, and redefine problems in an attempt to identify alternative strategies and solutions that might not be instantly apparent with the initial level of understanding. At the same time, Design Thinking provides a solution-based approach to solving problems. It is a way of thinking and working as well as a collection of hands-on methods [5]. According to Tim Brown of IDEO "Thinking like a designer can transform the way you develop products, services, processes – and even strategy [6]. Most popular application style includes an iterative loop of Empathize, Define, Ideate, Prototype, and Test. This fundamental philosophy lends to many variations essentially defining a learning path that begins with an understanding of the end beneficiary and ends with a workable solution.

In this paper, design thinking perspective is applied in a nested manner. First to design a management approach that begins with a survey to empathize with the NDE community and ends with an outline of a strategy around Why, What, and How of NDE 4.0; Second variation is within the adoption process of 'How' that defines a process from inspection insights to creative digitalization.

Figure 2 shows the outline of the design thinking philosophy as applied to NDE 4.0. The paper begins with an overview of the Industry 4.0 emerging technologies (Define) and a qualitative survey of stakeholder perceptions (Empathize). These are later used to develop various use cases of technology and related adoption challenges (Ideate). All of this leads up to the definition of a purpose and process for NDE 4.0 (Prototype of the Strategy). The actual execution is not perfectly linear and must be treated as iterative to keep it meaningful and dependable. As indicated by the feedback arrows and common in design thinking approach.

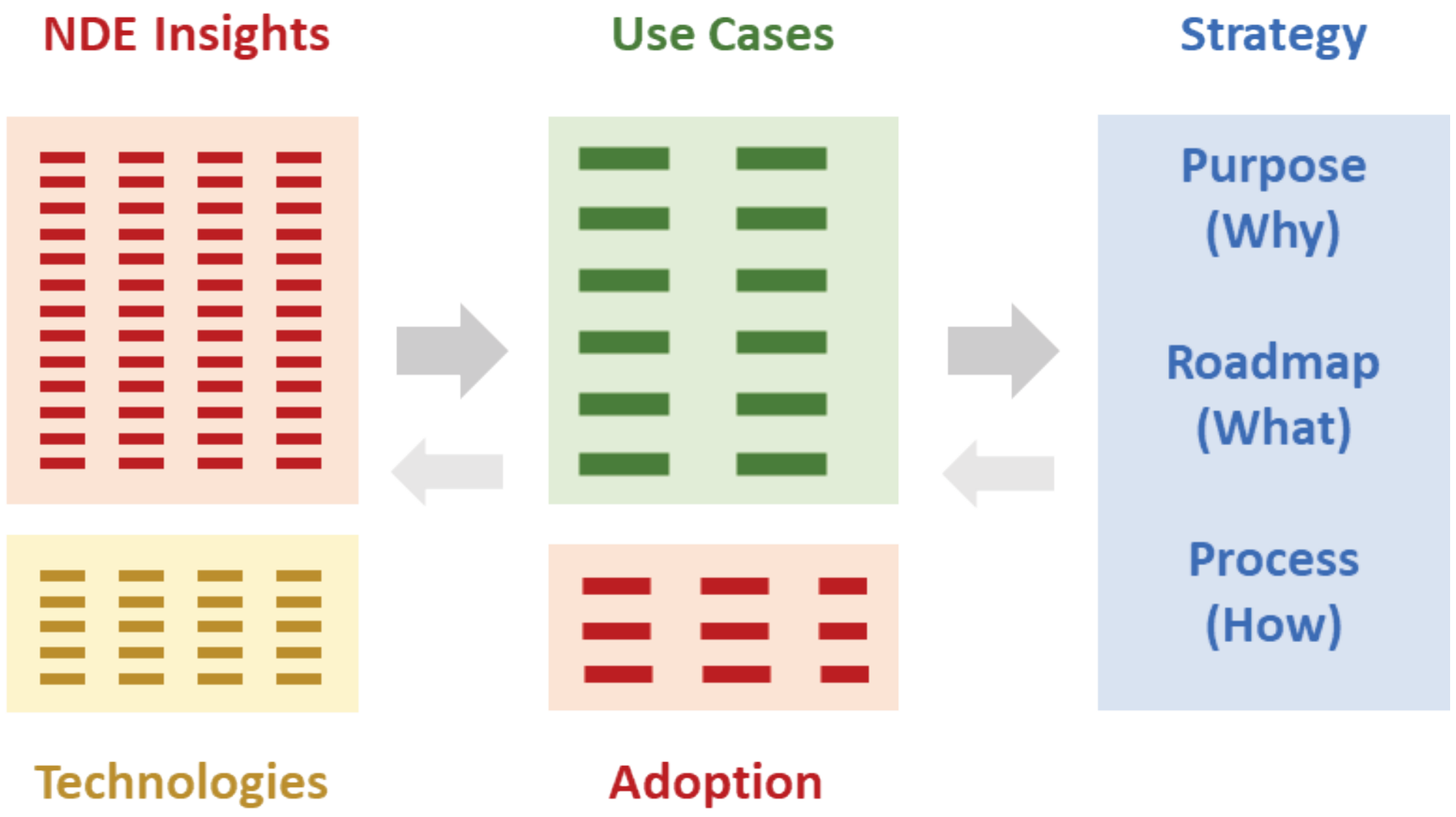

**Figure 2** Design thinking philosophy for developing the NDE 4.0 strategy. NDE Insights (Empathize), Technologies (Define), Use Cases and Adoption (Ideate), and Strategy (Prototype)

## *Technologies Driving the 4$^{th}$ Revolution*

Every thought leader has their favorite top 5 digital technologies as a part of industry 4.0. Which is more important depends upon the application. From the diversity typical of NDE domain, all of them will be discussed briefly. This enables a later discussion whether those technologies give more or less value to the outcome of NDE and if they even create new use cases.

This portfolio of technologies discussed begins with meaningful data collection (digital twin, digital thread, Industrial Internet of Things (IIoT), and semantic interoperability to enable machine readability). It continues with technologies enabling new ways of data transfer (5G), revision safe storage (blockchain), computing (Cloud, AI, Big Data, and Quantum Computers), and visualization (XR). Finally, multiple technologies (AM, Automation, Simulation, Reconstruction, and Digitization) enabling automation, data processing and purposeful application, are discussed in a new light with the fourth revolution.

### Digital Twin

A digital twin has to be treated as a concept. It provides value on three fronts:

1. It encapsulates all relevant data of an asset or a component (or respectively uses the data stored in multiple computer systems/databases using their semantic interoperability accessed by the IIoT.
2. It enables simulation of the asset usage based on the stored data.
3. It enables users to visualize the data and the simulation results.

For example, a digital twin of a human would encapsulate physical data like dimensions (weight, length, …), financial aspects, connections (friends, colleagues), eating and drinking preferences, health history, … etc. Social media cites (like Facebook or LinkedIn), user accounts (like Google or Apple), health insurance records or governmental records can be individually considered partial digital twins. With simulation tools added the behavior or the lifetime of a human can be predicted by those digital twins. This shows the value of data and importance of data security and data sovereignty.

One of the first implementations by an independent body for a digital twin is the asset administration shell of the platform Industry 4.0 [2,3].

Digital twins should be differentiated by the type of asset they represent. Those types include production facilities, production equipment, assembled products, components, inspection systems, devices, sensors, and operators. Digital twins can be layered and allow inheritance. For example, a digital twin of a production facility can contain the digital twins of all the production equipment and inspection systems within the production facility. And the digital twin of an inspection system can contain the digital twins of all sources, sensors, detectors, and manipulators.

Computer Aided Design (CAD) Systems could be viewed as a quite simple digital twin as they integrate the design data of a component, simulation, and visualization processes. However, it is firstly a digital twin of an early development state of a component as it does not contain, for example, any operational data and secondly it only contains the dimensional design data. This indicates that a complete digital twin might be difficult to achieve.

For NDT purposes two main digital twin types have to be differentiated:

1. the digital twin of the component to be inspected and
2. the digital twin of the inspection system / equipment.

The digital twin of the component to be inspected could contain product model, usage space, performance parameters, inspection, and maintenance records, etc. and will help the component owner to help improve production, design and maintenance. The digital twin of the inspection system helps to improve the inspection process.

**Digital Thread**

A digital thread, proposed by the US Department of Defense [7], connects the data from the planning and design phase of an asset over production and service until it goes out of service and allows to trace every decision and its implications. This means digitalization and traceability of a product "from cradle to grave". and it is more than a life cycle record as it also includes all the information from the planning and design phase. It can be seen as a successor to PLM (Product-Lifecycle-Management) enabling the realization of dream of the PLM systems.

**IIoT: Industrial Internet of Things and the Infrastructure**

The industrial internet of things (IIoT) connects assets with each other, with the digital twin, with the digital thread, with data-base systems, and with the cloud. For all those connections it uses open standard communication interfaces like OPC UA, WebServices, or oneM2M, and core gateways to link the core communication standards. The IIoT requires semantic interoperability. The development of each of the standard communication interfaces is driven by organizations such as OPC UA Foundation. The Industrial Internet Consortium (IIC) requires such core gateways in its standards and the International Data Spaces Association (IDSA) is in the process of implementing such gateways. The so-called IDS connectors connect all the communication standards with digital twins, the cloud, and data markets while guaranteeing the data sovereignty [2].

Similar concepts, like the building internet of things or the infrastructure internet of things, are conceivable for every industry and use case.

Within the inspection world, IIoT makes for example remote inspection, collaborative decision making, inspection workflows, archiving, and integrated inspection system design a reality. Moreover, it is the precondition to integrate NDE as one of the key data sources into Industry 4.0.

**Semantic Interoperability, Ontologies**

To effectively deal with communication standards, digital twins, digital threads, or data types/formats one key aspect is to give data unambiguous and shared meaning, which is achieved by semantic interoperability, and by ontologies. Syntactic interoperability, which is the necessary basis for semantic interoperability converting the data from a system-dependent to a system-independent format, enables data exchange between different systems. However, for a computer to understand data (just like for a human) the computer needs to know the unit of the submitted number, it needs to know whether the number refers to a height, a length, a weight, a time, … , and it needs to know the connections between the data objects. For example, that a car has a speedometer and the speedometer shows a value of 136 km/h. This is enabled by semantic interoperability by not only storing a value but also by identifying its connections and giving it a meaning.

Semantic interoperability converts data into information and enables different inspection equipment, asset digital twins, data analytical tools to communicate and understand each other, for a meaningful inspection outcome.

**Industry 4.0 Data Processing**

Some examples for Industry 4.0 data processing, also known as big data analysis, are digital engineering, feedback loops, trend prediction, probabilistic lifing, predictive maintenance, behavioral analytics, risk modelling or reliability engineering. All those tools are intended to improve design, production, maintenance, etc. and are based on data-derived knowledge. This requires the conversion of the data from IIoT, digital twin, digital thread, computer systems, databases, etc. into information by using semantic interoperability and the conversion of the information by statistical analysis or AI into knowledge.

**5G**

In layman language, the 5G is viewed as a successor to 4G for faster mobile data exchange. This part of 5G is called eMBB (Enhanced Mobile Broadband). However, it is the speed, range, and device density which makes 5G one of the corner stones of Industry 4.0. 5G brings Ultra-Reliable Low Latency Communications (URLLC) which allows robust real time data connections (latencies < 1 ms) and extended mobility (> 500 km/h). The Massive Machine Type Communications (mMTC) allows the connection of high density of devices (1 million/km$^2$) and cheap low complex mobile implementations.

5G provides the necessary bandwidth for high speed remote inspection and large-scale implementation – even for cheap inspection equipment – and could lead to the possibility for every inspection device to be accessed through IP connections.

**Blockchain**

Blockchains present a way to assure that data is not changed after being stored once. The security is enabled by a chain of data blocks, where every block consists not only of new data but also of a hash representing the data of the previous block (which in itself contains a hash of the block N – 2, and so on). This makes it difficult to change earlier blocks as the hashes in all the following blocks would have to be recalculated. To further enhance the manipulation safety, a '*nonce'* (number used once) is added to each block. A nonce is the result of a restricting rule for the hash of the following block. Meaning, there is a restricting rule for the hash of block N (for example it has to be smaller than a certain number) and, the nonce of block N-1 (which is included into the calculation of the hash of block N) has to be changed multiple times until the hash of block N fulfils the given rule. The technology emerged out of the need to secure financial transaction records or component files. For crypto-currencies, the rules for the nonce are usually increasingly difficult to achieve to ensure the value of the currency due to the difficulty of finding a fitting nonce.

In particular in quality assurance and maintenance it is important to guarantee that results obtained and reported cannot be changed in the future or that changes are tracked. This can be guaranteed by Blockchains.

**Cloud**

Another technology connected to Industry 4.0 is cloud computing, which enables the access to an IT infrastructure using the internet from any computing device around the globe and allows data storage, data processing and application software as a service. For a reasonable use of clouds semantic interoperability of data and of the communication interfaces are key so that the standard data processing and visualization software available in the cloud can understand the data.

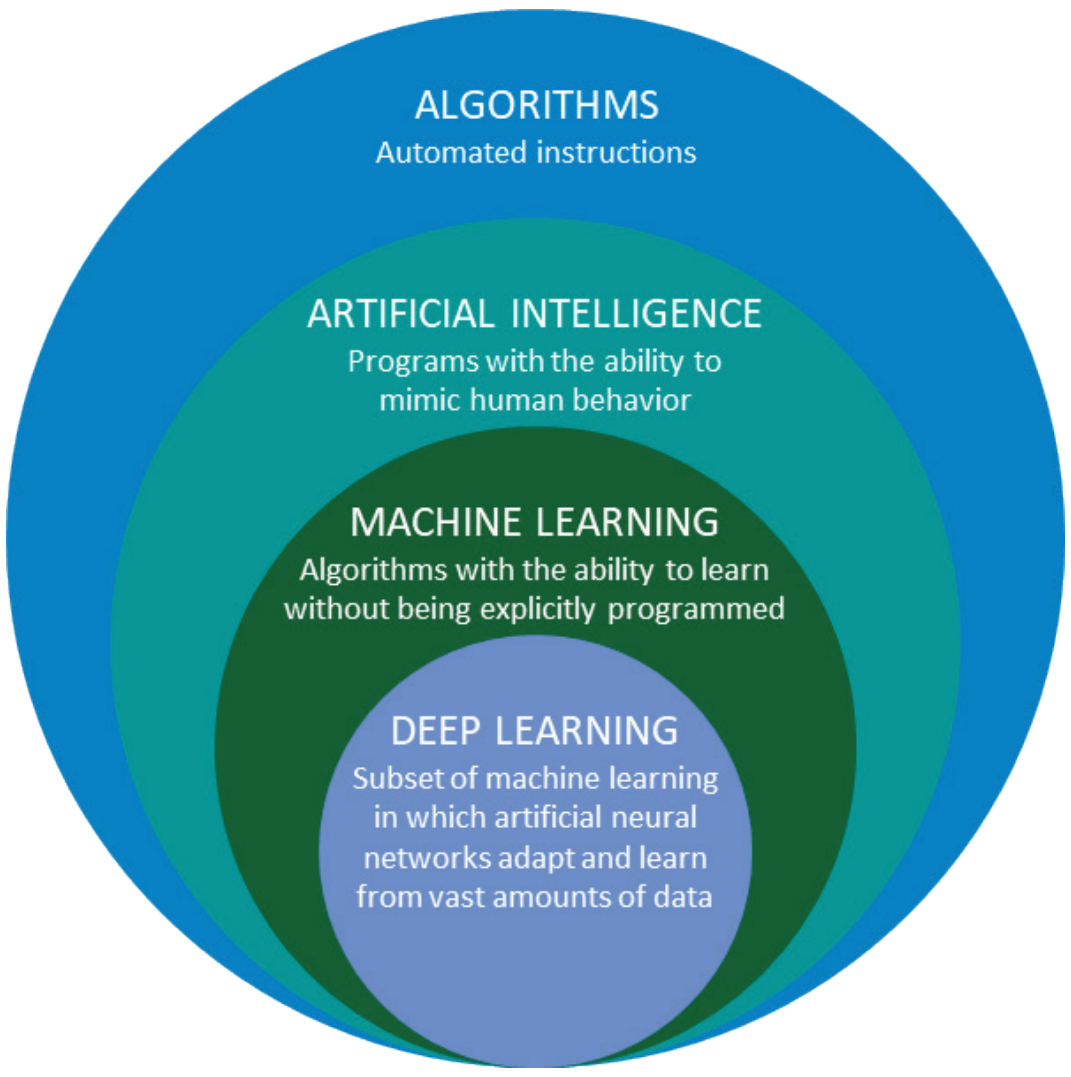

**Figure 3** Visualization of algorithms vs. artificial intelligence vs. machine learning vs. deep learning (Author: Johannes Vrana, Vrana GmbH, Licenses: CC BY-ND 4.0)

**Artificial Intelligence (AI)**

Artificial intelligence is defined as "A system's ability to correctly interpret external data, to learn from such data, and to use those learnings to achieve specific goals and tasks through flexible adaptation" [8]. In fact, as Figure 3 shows, artificial algorithms are a subset of algorithms which have the ability to mimic human behavior. Machine learning algorithms are AI algorithms with the ability to learn without being explicitly programmed. Deep learning algorithms are machine learning algorithms in which artificial neural networks adapt and learn from vast amounts of data. An intermediate step towards AI is intelligence augmentation where it helps the human with information and know-how to assist in decision making rather than replace the human judgement.

AI technology can help improve inspection reliability through comprehensive interpretation of inspection big data, not humanly possible. However, the implementation of AI requires a high skill set, both on the inspection process and the machine learning algorithms to ensure that system output is dependable. Otherwise, there is a risk of capturing and duplicating human incompetency. AI technology can also help to improve Industry 4.0 Data Processing.

**Big Data**

Big data is a field treating multiple challenges resulting from the vast amounts and different types of data collected by the IIoT using machine to machine and human computer interface. Those challenges include capturing, storing, analyzing, searching, sharing, transferring, visualizing the data which cannot be processed by traditional software.

The collection of data, and the conversion into information and knowledge is the main idea of Industry 4.0 Data Processing. With the collection of bigger amounts of data, the potential for enhanced knowledge is created which could lead to enhanced design, production, and maintenance. This shows the criticality of solving the big data challenges.

Within NDE, ability to manage big data can be very useful over the life span of an asset or the fleet, as the ways, forms, and volume of data continuously evolves.

**Quantum Computers**

Quantum computers use multiple entangled qubits (53 in [10]) to conduct calculations. The possibilities arising due to quantum computers are astounding but the working principle is hard to grasp. It is based on two quantum mechanical phenomena: superposition [11] and entanglement [12]. Those phenomena can only be observed in a microscopic world (~ size of an atom) and not in the macroscopic world. This is why they seem to even contradict our everyday experience, but they can be used to build computational devices working completely different than our current computers. Back in the early days, Schrödinger, one of the fathers of quantum mechanics, didn't accept the validity of superposition. This is why he wrote his paper which includes the famous thought experiment on the cat [11]. Eventually, Schrödinger had to accept that this thought experiment was misleading and accepted superposition.

Even that the working principle is hard to understand and beyond the scope of this paper, the possibilities of Quantum Computers are so immense that Quantum Computers could become the most important element to the fourth revolution. Unlike classical computers the results of all possible variations of input parameters are calculated at the same time. This enables an exponential speedup of certain algorithms compared to classical computers. Quantum computers will be more than just the next revolution within computing technology.

Quantum computers are now moving out of research to first applications. They will not replace classical computers, but they will be an incredibly powerful add-on for certain computational challenges. Currently quantum computing is not considered a prime Industry 4.0 technology, but this could change fast. In particular as quantum computers could play an essential role in solving some of the big data challenges and enabling game-changing artificial intelligence algorithms.

**Extended Reality (XR)**

Virtual Reality (VR) utilizes headsets to display a computer-generated 3D image (for example computer games), blocking the normal vision. Augmented Reality (AR) presents an interactive experience of a real-world environment enhanced by computer-generated image. Mixed Reality (MR) combines the VR, AR, and the physical world. Finally, the term XR (Extended Reality) is used to summarize all forms of digital enhancements to visualize data overlays in the physical world.

XR brings multiple improvement opportunities such as superior inspector training, real-time work instructions, ability to visualize anomalies as virtual overlay on real asset, visualize simulated future, and remote expertise consulting.

**Additive Manufacturing**

Additive manufacturing or 3D printing encompasses multiple new manufacturing technologies for various materials which add material layer by layer to shape a 3D object. This stays in contrast to classical subtractive manufacturing methods removing material from a block of raw material to achieve the shape. This has multiple benefits, in particular lot-size-one, creation of internal structures, embedded sensors, which would not be possible or immensely difficult to manufacture with subtractive manufacturing.

3D printing makes the nondestructive evaluation of such components challenging as discussed later. Moreover, for small lot sizes the state-of-the art procedures for reliability assessments need to be reconsidered.

**Automation, Robotics and Drones**

Automation, robots and drones evolved during the third revolution [2,3]. However, their ability to interpret and adapt to the environment makes them a part of the Industry 4.0. The ongoing activities could be considered enhanced automation, where the machine can choose to scan, and capture data based on observations rather than pre-programing.

For example, collaborative robots (cobot) allow a shared working space for humans and robots in production environments by assuring the safety of the humans by (for example) sensors. This makes the typical fences unnecessary and allows a direct interaction between a human and a robot.

Automation and AI technology can keep inspectors away from harm's way, by accessing the confined spaces, hard to reach areas, heights and depths, radiation exposure, and extreme weathers.

**Simulation and Reconstruction**

Simulation and reconstruction from the 3rd revolutions brings in a new meaning into the 4th by enabling real time control of physical actions based on predictive analysis. This provides both a capability and purpose to the digital twins and digital threads.

In the asset sustainment space, simulation technology enables optimization of the inspection program tied to parameters of choice – risk, cost, and downtime.

**Digitization, Digitalization, Digital Transformation, and Informatization**

Digitization is the transition from analog to digital and Digitalization is the process of using digitized information to simplify specific operations [2]. Digital transformation uses digital infrastructures and applications to exploit new business models and value-added chains (automated communication between different apps of different companies) and therefore requires a change of thought processes. Digital transformation requires collaboration for an improved digital customer experience. Informatization is the process by which information technologies, such as the World Wide Web and other communication technologies, have transformed economic and social relations to such an extent that cultural and economic barriers are minimized [9].

Digitization is the core of the third revolution, digitalization marks the process to the fourth revolution, and digital transformation is the core of the fourth revolution.

## *So, What is Industry 4.0?*

For all the technologies described above somebody might argue that none of those technologies is actually new and there is an element of truth to those statements. Industry 4.0 is not a single technology; it is a suite of cyber-physical technologies. The fourth revolution is not a discrete event; it is a phase over which the suite of cyber-physical technologies is coming together to change the way humans work and live, produce and consume, learn and stay healthy, and other things along the way like NDE.

So, **overall,** what makes Industry 4.0 is not the emerging technologies, but their integration for a purpose, that was not achievable up until now. The increase in communication bandwidth from 5G, ability to manage terabytes of data, computational processing speed and capacity, mobile devices, location services, ease of programming, have all enabled such integration. In the past strict customer retention was key for suppliers in all industries, but new business models are arising to enable the economic use of data by enabling collaboration. It is the reduction of burdens, the reduction of proprietary data formats and proprietary interfaces. It is the collaboration of different players around the globe to work on the greater good. This will eventually lead to a completely new market – a market for data – and a market for purposeful application of data.

## *And, What is NDE 4.0?*

Multiple papers have been published on NDE 4.0 [2,3,13,14,15,16,17] to bring awareness on this topic. They focus on single use cases and on comparisons across the four industrial revolutions and the four revolutions in NDE [4, 18,19]. In summary NDE 4.0 can be defined as "*A Cyber-physical Non-Destructive Evaluation (including testing)*; arising out of a confluence of industry 4.0 digital technologies, physical inspection methods, and business models; to enhance inspection performance, integrity engineering, and decision making for safety, sustainability, and quality assurance, as well as provide relevant data to improve design, production, and maintenance through useful life."

### The Evolution as seen by the Authors

For centuries humans have taken care of their safety using the five basic senses – touch, sight, hearing, smell, and taste. Over the last 200 years this has evolved into a planned and instrumented approach to meet the needs of the industrial revolutions [2]. For the definition of the four NDE revolutions it was chosen to define them independent from Industry and by the revolutionary changes within NDE. This results in the fact that from strict timetable view the industrial revolutions and the revolutions in NDE did not occur at the same time (in most cases the NDE revolutions are belated), but as shown in Figure 1 the commonality between the Industrial and NDE revolutions are getting stronger with each revolution.

**NDE 1.0:** In sync with the 1st industrial revolution of steam powered systems, came the need for tracking safety assurance through periodic checks using a prescribed process, and perhaps simple tools to enhance human capability. Touch enhanced with use of a tapping device, vision enhanced with magnifying glass, or applying color, etc.

**NDE 2.0:** The 2nd industrial revolution marked by electricity, brought analog devices, to address the need to look beyond the line of sight, using waves and frequencies outside human sensory range such as ultrasonic, eddy currents, and x-ray. Simultaneously we saw the emergence of enhanced visual techniques like fluorescent dye penetrant and magnetic particle, to reveal surface anomalies.

**NDE 3.0:** The 3rd industrial revolution through digital technologies offered another leap in managing inspection data acquisition, storage, processing, 2D and 3D imaging, interpretation, and communication, including programmed movement of sensors. For example, digital visual and infrared sensor arrays for cameras, digital UT, ET, RT, and computed tomography.

**NDE 4.0:** The 4th industrial revolution integrates the digital tools (from 3rd) and physical methods of interrogating materials (from 2nd) in a closed loop manner transforming human intervention and enhancing inspection performance. Within the context of the physical-digital-physical loop of NDE 4.0; digital technologies and physical methods may continue to evolve independently, interdependently, or concurrently. The real value is in concurrent design of an inspection through application of Digital Twins and Digital Threads. This provides ability to capture and leverage data right from materials and manufacturing process to usage and in-service maintenance. The data capture across multiple assets, can be used to optimize prescriptive maintenance, repairs and overhauls over the life time of an asset. The relevant data can be fed back to OEM for design improvements.

NDE 4.0 also serves the emerging trends in custom manufacturing. Remote NDE can keep the inspector away from the harm's way and integration of "tele-presence" can bring additional specialist in the decision process from anywhere in the world quickly and affordably.

Figure 1 summarizes the revolutions of NDE and Industry. One of the main differences in the definition of the NDE revolutions compared to some of the existing and beforementioned publications is that human perception is considered as the basis for the 1st NDE revolution. This goes along with the definition of handcraft as the basis for the 1st industrial revolution.

A new revolution in NDE does not imply that work on the previous ones should be stopped. A new physical method created today does not automatically qualify for NDE 4.0. For example, a better fluorescent dye penetrant discovered today is still NDE 2.0. It could be within NDE 3.0 when applied and/or photographed by a robotic arm; and within NDE 4.0 if the photograph is embedded into the digital thread and/or interpreted using AI along with similar pictures stored in the digital thread, for dependable decision making.

## *NDE and NDE 4.0 Insights for Design Thinking*

Design Thinking revolves around a deep interest in developing an understanding of the people for whom the products or services are designed. It helps to observe and develop empathy with the target user. So, engagement needs to be started with practitioners across the NDE value stream, with surveys as well as interactive engagements at training workshops, and virtual sessions.

In this paper two of the surveys completed so far are discussed. The first question "What is NDE 4.0?" is just like the question "What is Industry 4.0?", not easy to answer and different people have different views. And that is what the authors were looking for. The second question "what are your current NDE challenges" became a catch all diverse aspects again offering an opportunity to define what NDE 4.0 should be. To discover more, some research on industry 4.0 adoption was performed. The results presented below and in table 1 are used to reveal different use cases for NDE 4.0 in the subsequent sections.

### *Survey 1: What do you think is NDE 4.0?*

This survey was conducted at the short course for NDE 4.0 during the ASNT annual conference 2019 before starting the 4-hour training, with 60+ in-person participants. The interpretation was further enhanced through interactive exchange during the course to credible summaries. The bullet points in the following six sections represent the answers to the survey. It was chosen to consciously share raw data, at the cost of excessive detail, because the need to record and share the emotional state of the practitioners is crucial, before summarizing the interpretation.

**NDE 4.0: The Future of NDE**

- *"The next level."*
- *"It's a good future for the NDT work."*
- *"The next generation of NDE leading with data, automation, and digitization."*
- *"Think it is just getting started but has potential."*

**Remarks**: The participants realize and identify NDE 4.0 as the future of NDE.

**NDE 4.0: Enhancing NDE Methods by Emerging Technologies**

- *"Information/Communication technologies coupled with augmented inspection methods."*
- *"A means to maintain and advance our craft utilizing a melding of best practices and emerging technologies."*
- *"IA, XR, Deep Learning."*
- *"IoT, Big Data, networking, etc.: Taking these components and applying them throughout the NDT industry."*
- *"Deep Learning."*
- *"Improving NDT efficiencies via modern connectivity and advanced computing."*
- *"Utilization of technical and social advancements. As technology increases so does the capabilities/possibilities. As networking, data collection/analysis/reporting, automation increase so should the influence on NDE."*

**Remarks:** Participants correctly understood the use of the emerging technologies portfolio, described previously, to enhance existing NDE methods. This could also be called "Industry 4.0 for NDE".

**NDE 4.0: Data source for Industry**

- *"If we can use NDE results and feedback to improve quality of products."*
- *"Acquisition, manipulation, and interpretation of data obtained utilizing NDT techniques to determine the effectiveness of a manufacturing process or application."*
- *"To increase quality and production."*

**Remarks:** NDE has always been a central tool to assure a sufficient quality and safety during industrial production and operations. The respondents appreciate the potential for an increased use of the results of NDE to improve the design, the product, and the production by statistical analysis of the NDE data. This could also be called "NDE 4.0 for Industry" and plays along with the wish to give NDE data more meaning.

**NDE 4.0: Giving more meaning to NDE data**

- *"The conglomeration and identification of data for use. Giving meaning."*

**Remarks:** Single response should be considered limited input. However, it indicates that participants are just beginning to appreciate the idea that NDE 4.0 should give more meaning to NDE data. Based on subsequent engagement, the answer can be seen in two ways. First, as discussed above, by using NDE data for statistical evaluation in an Industry 4.0 environment, and second, for inspection control through digital workflows, traceability, or reliability.

**NDE 4.0: Data Exchange and Feedback**

- *"Feedback, data access to clients for production; EOL assessment of equipment."*

**Remarks** – The single answer goes along with the previous idea; however, it adds two additional key points: data exchange between customer and client – and an End of Life (EOL) assessment with feedback to the original manufacturer. The first point shows the desire to enable the electronic exchange of job descriptions and inspection results (using standard interfaces) and the second the desire to learn more about the development of equipment during its life. Both for giving NDE data more meaning.

**NDE 4.0: Human Implications**

- *"Eventual robots interpreting better than humans."*
- *"Human with AI will replace human without using AI."*

**Remarks:** The last group of answers show the human side of adoption (or implicit resistance and fear). The human side of NDE 4.0 comprises a whole new set of elements, such as mental health of inspectors, career development or changes in training and certification models that will be discussed in future publications of the active participants in the NDE 4.0 community. Participants have not yet fully conceived this side due to lack of direct experience.

## *Survey 2: What are your challenges in NDE?*

This survey pulled data from two sources – the course on NDE 4.0 discussed earlier, and an open-ended engagement on social media [2,3]. Since the why and what of NDE 4.0 are not yet broadly accepted in any industry, the question was inversed to state "What was the most negative thing somebody said, or what are your most negative thoughts about NDT/NDE ?" Once again, the bullet points reflect raw data under the following fourteen topics sections, with some editorial comments indicated by brackets.

The summaries provide a list of challenges NDE community must take on. Some of those challenges can be addressed with NDE 4.0; conversely NDE 4.0 can be designed to resolve what matters. This is Design Thinking approach – a creative problem-solving process focused on humans in the loop, which leads to better products, services, and internal processes.

**NDE: Not a Skilled Trade**

- *" "NDE is not a skilled trade" is something I've heard over and over by some men in "skilled trades" ".*
- *"We don't need NDT - you only test [and introduce] flaws into the material."*

**Remarks:** These show the perception of other skilled professionals regarding NDE, who do not consider it to be a skilled trade and on top some people incorrectly assume that NDE is causing flaws. Both answers might be caused by the fact that the awareness of NDE is quite limited in society and work places. NDE 4.0 is likely to show value through data. To raise awareness the NDE community needs a proper campaign around its value proposition to everybody - industry, governance, society, education, and media, even before adoption can be expected.

**NDE: Unnecessary Cost Factor**

- *"NDT is all smoke and mirrors".*

- *“Production brake".*
- *“Turnover preventors".*
- *“Unnecessary cost factor".*
- *“Non value added” [Multiple responses].*
- *“You are like my mother in law, I don't need you... hate it when you are there... you create extra work for the rest of us and I end up paying a ****load of money".*
- *“My negative thoughts center on my company’s lack of understanding of NDT and the fact that they see it as an impediment to throughput".*
- *“Costs money and takes time".*
- *“Expensive to use and only slows down manufacturing".*
- *“NDT is a detriment to production more than an additive value".*
- *“Costs too much".*
- *“It's all smoke and mirrors; costs too much; bottle neck; non-value-added; only represents negative issues".*
- *“Some have said it is unnecessary and has no effect on the product or companies’ future".*
- *“If NDT becomes mandatory, our product will be too expensive for the market”.*
- *“NDT does not have any value at all. It only sorts out parts, that in reality are good. I don't want it and I would never ever do it, but my customer insists on it. I'd prefer spending the money into further improvement of my production!”*

**Remarks:** Overwhelming response is to the point of frustration that others do not see NDE / NDT as a value add, but a cost intensive activity and a production break; up to the point that it would make the product too expensive for the market. Opportunity appears to be in improving production. And that can be done through data where, “NDE meets IIoT” as identified in Survey #1 leading up to “NDE 4.0 for Industry”.

**NDE: NDE in Engineering**

- *“Many times, other Engineers and project managers never include NDT Engineering in planning because they believe they know everything there is to know about NDT. Many times, mindlessly prescribing methods that cannot detect the flaws or just throwing it in after planning with even thinking. NDT Level IIIs and Engineers should always be included in design and planning phases. This will save money on the long run.”*
- *“We don’t need NDT, the safety factors in design (civil engineering) will cover any flaws (and probabilities will cover any uncertainties)”*
- *“your xxx is not good enough to detect yyy, so I won't use it for zzz (look for the logic)”*

**Remarks:** In some cases, NDE engineering is not included in the design phase of products either because other engineers think they have a sufficient knowledge of NDE or because they think that NDE is not needed due to safety margins. NDE 4.0 for Industry could also help here. If NDE data is seen as valuable, the focus might shift to maintain the data treasure produced by NDE and the statistical evaluation is used to decide on appropriate safety factors if the inspections are to be waived.

**NDE: Hard to Explain**

- *“It is sometimes hard to explain the benefits to non NDE people.”*
- *“Technology is too Star-Treky.”*
- *“For UT & ET: Still black magic.”*
- *“My negative thought would be that NDT needs better distribution of education and understanding to the public.”*
- *“Disbelief that UT is finding defects.”*

**Remarks:** If non-NDE engineers would know enough they would easily see that the specialist knowledge of an NDE engineer is key. It needs a fundamental understanding of the inspection technology, of the physics and of the detection mechanism. This also leads to the fact that some people are quite intimidated by NDE or even don’t believe in the results. NDE 4.0 might aggravate this situation, if not properly understood.

**NDE: Human Errors**

- *“Too many operator variables.”*
- *“Based on operator guess.”*
- *“Operator dependent.”*
- *“Consistency between inspectors to interpret the signals.”*
- *“Human error included in process.”*

- *“The most negative thing I have heard is there is too much variance potential in interpretation of results from person to person. This causes question of the reliability of the results. I agree with the statement.”*
- *“Human factors are very important in risk-based management.”*

**Remarks:** The high number of parameters and the low level of automation leads to highly complex inspection situations for the inspectors. Therefore, human factors have to be seen critical for NDE. NDE 4.0 will help to reshape human intervention by increasing the automation level of inspection and data interpretation and assuring revision-safe storage.

**NDE: Reliability**

- *“Only finds ghosts.”*
- *“Difficult to eliminate false positives.”*
- *“Find unnecessary defects; not proven.”*
- *“Didn't predict a failure.”*
- *“Risk outcomes for miss-calls in NDE are higher, making it more responsible and skill critical field whether its Aerospace, pipeline, or refinery work.”*

**Remarks:** Inspection system reliability consists of the reliability of the method mixed with the human factors and application factors. And the proper determination of the reliability, using PoD (Probability of Detection) or IoU (Intersection over Union) considerations, will get even more crucial once NDE data is being used for Industry 4.0 data processing, like probabilistic fracture mechanics [20].

**NDE: Lots of Data**

- *“A lot of information to analyze.”*

**Remarks:** What will be a huge benefit for Industry 4.0 in the future is currently a challenge for NDE: some NDE methods create file sizes of several gigabytes which leads to the fact that NDE inspectors have to analyze lots of data. Technology advancements in storage and processing hardware paired with reconstruction, AI, and solving some of the big data challenges will ease the information overload and will contribute with accelerated knowledge and insights creation.

**NDE: Very Conservative**

- *“My negative thoughts for NDE are that compared to other fields such as medical, automotive, etc. it is not on par in advancing as quickly as technological advances become available.”*

**Remarks:** The NDE business must assure the quality of assets. Therefore, the trust in the inspections needs to be build and finally the best-practices need to be standardized. This are some of the reasons why the NDE business is conservative and resists innovation. This will also be a challenge for the introduction of NDE 4.0. However, NDE 4.0 could also be the means, by collecting and analyzing the data, to show easier that a new technology earned its trust. Increased safety and value creation are at the core of NDE 4.0.

**NDE: Methods and Technology Access**

- *“Do not always have technology to find what we are looking for.”*
- *“Operational hardship in military field application.”*
- *“Reference [knowledge base or access] is not up to the mark.”*

**Remarks:** There are cases where the technology, method or skill is not accessible. NDE 4.0 may not be able to resolve this challenge.

**NDE: Inspector Training on Manufacturing Methods**

- *“Lack of process knowledge and surface preparation.”*
- *“So many NDT inspectors who have not enough experience and little knowledge of welding making false calls”*

**Remarks:** To correctly interpret the NDE data, the inspector needs a basic knowledge of the manufacturing method of the asset under examination and of the key manufacturing processes causing the potential material imperfections. The XR references and AI algorithms should prove helpful to the inspector here. The inspectors with AI will replace inspectors without AI.

**NDE: Customers Influencing/Questioning the Results**

- *“Refused to believe or understand what test was telling them. Test not valid.”*
- *“Why don't you inspect at a different location?”*
- *“Perform the spot test at a different location.”*
- *“You mustn't look for indications in area you expect defects.”*
- *“You can use another method, then the findings are acceptable.”*
- *“I got the "the other inspector never rejected anything, why are you rejecting so many pieces" guess something in the process changed is what I said.”*
- *“The amount of welders who somehow think you have a magic pen for putting defects in radiographs is astounding. "That wasn't there when I welded it", says the welder with the X-ray vision!”*
- *“You don't need any inspection until something goes bang. Always chuckle when a welder tells you that they have never had a weld rejected. Two types of welders out there. There's those that accept that there's always a chance a weld will dip and there's those that tell a lot of lies.”*

**Remarks:** A proper inspection is key to guarantee the quality necessary for an asset. However, with NDE being seen as un unnecessary cost factor and with non-NDE personnel not sufficiently understanding the value of NDE it can come to situations that customers start questioning the results of the inspections or even that they start trying to influence the inspectors. NDE 4.0 will help through data-based evidence and XR enabled communication, showing the need and the success of NDE. The use of perfected human computer interfaces (HCI) through UX (User Experience) improved design will have a profound impact in NDE 4.0 applications adoption by business decision makers.

**NDE: Traceability**

Surprisingly, none of the answers was regarding inspection traceability

**Remarks:** This might be one of the major challenges of current NDE and NDE 4.0 should have a positive impact – ensuring that the correct component was inspected, the results are connected with the appropriate component, stored in a revision-safe manner, and are easy to retrieve in a digital manner.

This is an example of uncovering unknowns that people are unaware of. Or a new value proposition of NDE 4.0 never leveraged before.

**NDE: Compensation**

- *“Compensation tends to be dynamic per region versus static per method/experience/level/market availability of those methods (i.e. not a lot of certified ET III tubing inspectors).”*

**Remarks** - Job compensation, in particular in areas requiring responsibility, needs to be adequate to prevent persuasion or even bribes. Moreover, it seems to be an issue that the regional differences are dominant. NDE 4.0 might aggravate the situation by virtue of additional skill set required.

**NDE: Ethics / Inspector Mentality**

- *“Got offered something off the breakfast menu at McDonald's for me and my helper once on a turnaround. It was insulting because it was the ugliest weld I had ever seen on a 18in pipe and it was to 31.3 Severe Cycle too. It wouldn't have been as insulting had I been offered the dinner menu at least... Either way they had to cut it out cuz I don't do bribes.”*
- *“Lack of ethics in certification/qualification/training of technicians, and in the application of test procedures.”*
- *“I don’t inspect chips.”*
- *“Why do we need to inspect parts when we never find indications?”*

**Remarks:** The next step after simple influencing of inspectors might be bribes. This can bring inspectors into ethically difficult situations. However, also some inspectors / employers are trying to take shortcuts in training, certification and inspection procedures. The tendency of humans and inspectors to not perform tasks which seem unnecessary to the individual can be critical to safety or manufacturing quality. NDE 4.0, by integrating technologies like blockchain and enhanced automated data analysis may limit the possibilities for fraudulent and unethical behavior and can be quite a policeman by tracking the inspections, recording, and visualizing the inspections performed.

## *NDE 4.0 Use Cases and Value Proposition*

Stepping back and looking at the outcome of the two surveys, it can clearly be seen that the digital technologies can address several negative perceptions and key performance challenges of NDE. In the spirit of design thinking, various value propositions or use cases can be classified in two broad categories, (A) Industry 4.0 for NDE, and (B) NDE for Industry 4.0. Table 1 shows an overview of the different use cases and their connection to the challenges. The following presents an explicit look at them, as a synthesis of various implicit remarks from the surveys and research on this subject; eventually building up to the purpose of NDE 4.0.

| NDE 4.0 Use Cases | | | Survey 1: What is NDE 4.0? | | | | | | Survey 2: NDE Challanges | | | | | | | | | | | | | | NDE 4.0 Adoption Challanges | | | | | | |
|---|---|---|---|---|---|---|---|---|---|---|---|---|---|---|---|---|---|---|---|---|---|---|---|---|---|---|---|---|---|
| | | | NDE 4.0: The Future of NDE | NDE 4.0: Enhancing NDE Methods by Emerging Technologies | NDE 4.0: NDE – data source for Industry | NDE 4.0: Giving more meaning to NDE data | NDE 4.0: Data Exchange and Feedback | NDE 4.0: Human Implications | NDE: Not a Skilled Trade | NDE: Unnecessary Cost Factor | NDE: NDE in Engineering | NDE: Hard to Explain | NDE: Human Errors | NDE: Reliability | NDE: Lots of Data | NDE: Very Conservative | NDE: Methods and Technology Access | NDE: Inspector Training Regarding Manufacturing Methods | NDE: Customers Influencing/Questioning the Results | NDE: Traceability | NDE: Compensation | NDE: Ethics / Inspector Mentality | Existing NDE Standards | Digital Tranformation of NDE | Standardization | Validation | Leadership | Skills and Competenvies | Management Processes |
| Industry 4.0 for NDE | Emerging technologies | Digital Twin & Thread | X | X | | | | X | | | | | | | | C | | | | | | | C | | | | C | C | C |
| | | 5G | X | X | | | | X | | | | | | | | C | | | | | | | C | | | | C | C | C |
| | | Blockchain | X | X | | | | X | | | | | | | | C | | | | X | | | C | | X | C | C | C | C |
| | | Cloud computing | X | X | | | | X | | | | | | | | C | | | | | | | C | | | | C | C | C |
| | | Artificial intelligence | X | X | | | | X | | | | C | X | X | X | C | | X | | | | | C | | X | C | C | C | C |
| | | Big Data | X | X | | | | X | | | | | | | X | C | | | | X | | | C | | | | C | C | C |
| | | Quantum computers | X | X | | | | X | | | | C | X | X | X | C | | | | | | | C | | | C | C | C | C |
| | | Extended reality | X | X | | | | X | | | | X | | | | C | | | | | | | C | | X | | C | C | C |
| | | Additive manufacturing | X | X | | | | X | | | | | | | | C | | | | | | | C | | | | C | C | C |
| | | Robotics | X | X | | | | X | | | | | | X | | C | | | | | | | C | | | C | C | C | C |
| | | Simulation | X | X | | | | X | | | | | | X | X | C | | X | | | | | C | | X | C | C | C | C |
| | | Reconstruction | X | X | | | | X | | | | X | X | X | X | C | | X | | | | | C | | X | C | C | C | C |
| | | Digitalization | X | X | | | | X | | | | | | | | C | | | | | | | C | | | | C | C | C |
| | Inspection control | Traceability | X | | | X | X | X | | | | | X | | | | | | X | X | | X | C | C | | | C | | C |
| | | Digital Workflows | X | | | X | X | X | | | | | | | | | | | | | | | C | C | | | C | | C |
| | | Digital Commisioning | X | | | X | X | X | | | | | | | | X | | | | | | | C | C | | | C | | C |
| | | Feedback loops | X | | | X | X | X | | | | | | | | X | | | X | | | X | C | C | | | C | | C |
| | | Result Statistics | X | | | | | X | | | X | X | X | X | | | | | X | | | X | C | C | | | C | | C |
| | | Reliability | X | | | X | | X | | | X | X | X | X | | | | | | | | | C | C | | | C | | C |
| | | Usage statistics | X | | | X | | X | | | | | X | X | | | | | X | | | X | C | C | | | C | | C |
| | | Training status | X | | | X | | X | | | | | X | X | | | | | | | | | C | C | | | C | | C |
| | | Value of NDE | X | | | X | X | X | | X | X | X | X | X | | | | | X | | | X | C | C | | | C | | C |
| | Improving NDE equipment | Statistical evaluation | X | | | | | | | | | | | | | | | | | | | | | | | | C | C | C |
| | | Feedback loops | X | | | | | | | | | | | | | | | | | | | | | | | | C | C | C |
| | Remote access by | Equipment manufacturers | X | | | | X | X | | | | | | | | | | | | | | | C | | | | C | C | C |
| | | Experienced personnel | X | | | | X | X | | | X | | X | | | | X | X | | | | | C | | | | C | C | C |
| NDE for Industry 4.0 | Data transfer and storage | NDE for IIoT / Digital Twin | X | | X | X | X | | | X | X | | | | | | | | | | | | C | C | X | | C | C | C |
| | | Standardized Interfaces | X | | X | X | X | | | X | | | | | | | | | | | | | C | C | X | | C | | C |
| | | Semantic interoperability | X | | X | X | | | | X | | | | | | | | | | | | | C | C | X | | C | | C |
| | | Data security | X | | X | | C | | | | C | | | | | | | | | C | | | | | X | | C | | C |
| | | Data sovereignty | X | | X | | C | | | | | | | | | X | | | | | | | | | X | | C | | C |
| | | 5G | X | | X | | | | | X | | | | | | | | | | | | | C | | | | C | | C |
| | | Clouds | X | | X | | | | | | | | | | | | | | | | | | C | | | | C | | C |
| | NDE of | Additive manufacturing | X | | | | X | | | | | | | C | | | | | | | | | C | | X | | C | | C |
| | | Drones, critical robots | X | | | | | | | | | | | | | | | | | | | | C | | | | C | | C |
| | Data evaluation | Feedback loops | X | | X | X | X | | | X | | | C | C | | | C | | C | | | C | C | C | C | | C | | C |
| | | Manufacturing Optimization | X | | X | X | | | | X | | | C | C | | | C | | C | | | C | C | C | C | | C | | C |
| | | Trending | X | | X | X | X | | | X | | | C | C | | | C | | C | | | C | C | C | C | | C | | C |
| | | Digital Engineering | X | | X | X | | | | X | X | | C | C | | | C | | C | | | C | C | C | C | | C | | C |
| | | Probabilistic lifing | X | | X | X | | | | X | X | | C | C | X | | C | | C | | | C | C | C | C | | C | | C |
| | | Predictive maintenance | X | | X | X | | | | X | | | C | C | | | C | | C | | | C | C | C | C | | C | | C |
| | | Behavioral analytics | X | | X | X | | | | X | | | C | C | | | C | | C | | | C | C | C | C | | C | | C |
| | | Risk Modelling | X | | X | X | | | | X | | | C | C | | | C | | C | | | C | C | C | C | | C | | C |
| | | Reliability engineering | X | | X | X | | | | X | X | | C | C | | | C | | C | | | C | C | C | C | | C | | C |
| | Data monetization | New Business Models | X | | X | X | | | | X | | | C | C | | | C | | C | | | C | C | C | C | | C | | C |
| Awareness | Awareness | Awareness about NDE | X | | | | | | X | X | X | | | | | | | | X | | | X | C | C | | | C | C | C |
| | | Value of NDE | X | | X | | | | X | X | X | | | | | | | | X | | | X | C | C | | | C | C | C |

**Table 1** The cross correlation between the use cases identified and both the answers on the surveys “What do you think is NDE 4.0” and “What are your challenges in NDE?” and the NDE 4.0 adoption challenges. Here “X” symbolizes the cross correlation and “C” that a certain challenge might be critical for a use case.

## *Industry 4.0 for NDE*

### Enhancing NDE Capability and Reliability through Emerging Technologies

To begin with, most digital systems offer a clear advantage over traditional system in terms of accuracy and speed. However, a significant contribution of Cyber-Physical NDE system (NDE 4.0) stems from the better control or partial elimination of human factors in terms of reliability (PoD). Having developed techniques to quantify human factors in NDE-POD studies [21,22], author has a first-hand appreciation of value NDE 4.0 in context of system performance. This leads to a more reliable inspection system, i.e. better Probability of Detection (POD) and a more consistent POD from inspection to inspection. Ref Figure 4. Virkkunen et. al. [23] have shown that using sophisticated data augmentation, modern deep learning networks can be trained to achieve superhuman performance by significant margin in ultrasonic inspections.

All this adds to a dependable NDE permitting optimization of inspection program, saving time, money, and improving asset availability.

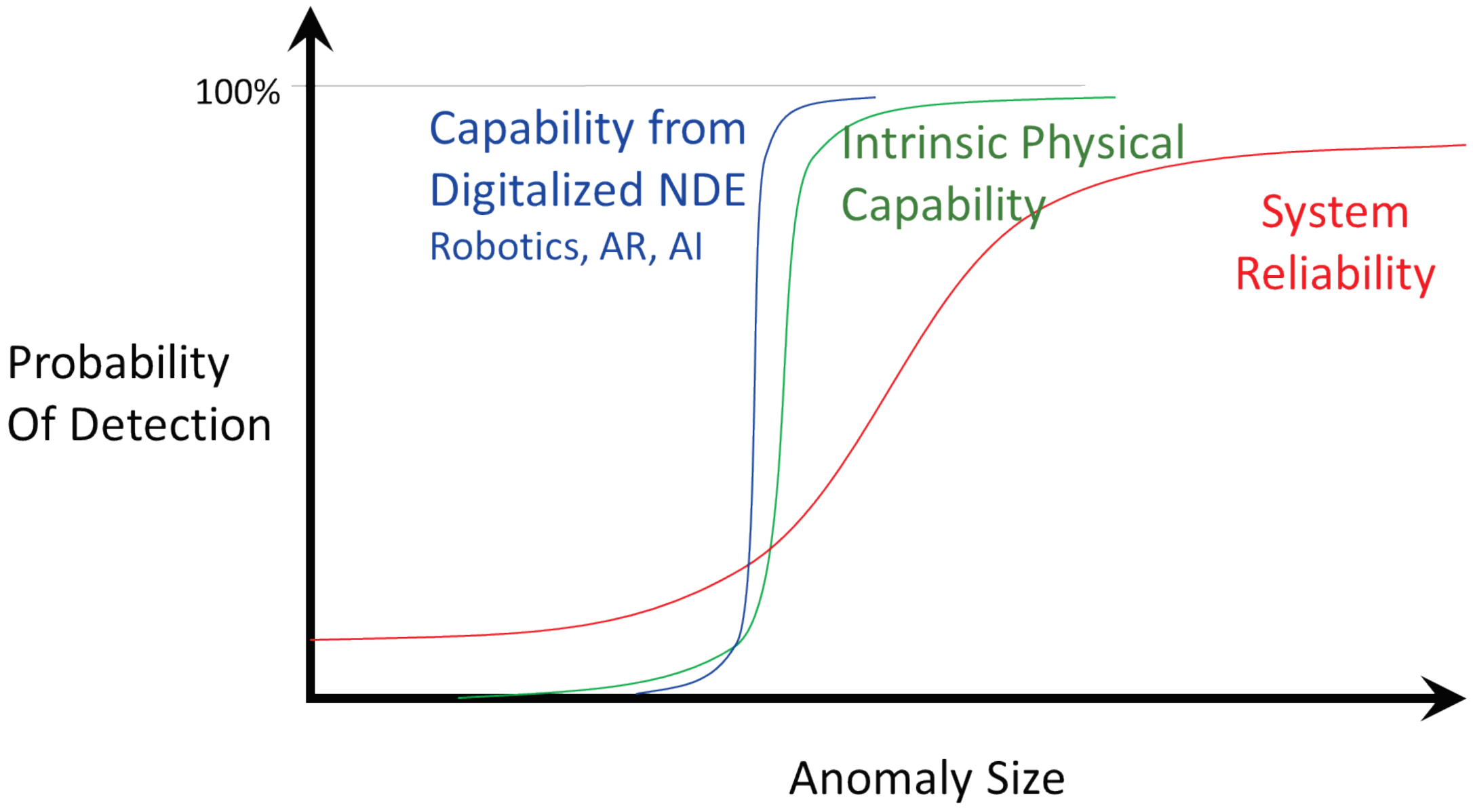

**Figure 4** Industry 4.0 Capability has the potential to provide a steeper asymptote to a POD of 1.0 as compared to Intrinsic Capability or System Reliability [16]

### Improving Efficiency and Effectiveness of Inspections through better Control

The entire inspection process can be made more effective and efficient through use of digital workflow control and tracking, just like so many manufacturing processes on shop floor or logistics in retail distribution. Multiple aspects have been identified.

Component traceability requires the need to ensure that the correct component was inspected, the documentation is revision-safely stored, and the results can easily be retrieved. Revision-safe data storage can be implemented by using blockchains and the component identification by digital component files and electronic component identifiers.

Digital workflows can enable full value stream efficiency: task allocation from the customer to the inspector and the results transfer back to the customer. Those data transfers are performed using IIoT technologies and interfaces (instead of Excel files or PDFs using email typical of 3rd revolution). Digital commissioning goes one step further by also transferring the order-related information using standard IIoT interfaces. Finally, with the implementation of digital supply chain processes both to customers and suppliers using standard interfaces a complete digital workflow can be established enabling NDE Processes 4.0.

NDE result statistics from production and in-service inspections, feedback loops from destructive tests, component acceptance, and EOL component testing, help quality assurance personnel to get a better appreciation of the value of the inspections. This data can also provide insight into system performance and reliability.

Usage statistics or inspection performance evaluations can show the need for a certain inspection on one hand and identify human factor influence on the other hand. This should help reduce operator dependence, inspection inconsistencies, and need for additional training or process change. Such evaluations can also be used to monitor training, experience hours for personnel evaluation, qualification, and certification. If required and permitted by local law, it can even indicate the mental state of inspectors, to provide support elements to improve their conditions such as stress monitoring or fatigue.

Overall, a better inspection control will help prove and visualize the value of NDE, helping the industry.

**Improving NDE Equipment using Inspector and Built-in Application Feedback**

The use case to improve NDE equipment using feedback wasn't mentioned by any of the answers of the surveys and from several personal conversations it seems that the NDE hardware manufacturers and software developers do not yet see a big value within this use case. The idea is simple: provide the data like error codes, system parameters, system status information, software exceptions/errors, or use or misuse back to the NDE equipment developers, so they can improve the equipment, the systems, or the software.

The statistical evaluation of user behavior helps improve the user interface design, training, and applicability. This feedback loop may also contribute to accelerated troubleshooting and improvements of inspection equipment as a competitive advantage.

**Improving Inspector Safety and Inspection Support through Remote Access**

Remote Support or TeleNDE by equipment manufacturers or other experienced inspectors/engineers can mitigate several challenges. Such remote support solutions or remote-controlled robot / drone-based equipment can be enabled by extended reality platforms and connected devices.

Remote Support by equipment manufacturers at the inspection location, in particular at hard to reach locations, could be a huge help to the inspectors and a money saver for the inspection service providers. At the same time, it is an opportunity for the manufacturer to investigate/reproduce potential issues/bugs with the equipment, if any.

Remote Support by other inspectors can help in inspection situations were a second opinion is needed, were an in-depth evaluation of indications identified by the inspector at location needs to be conducted and where local (potentially inexperienced) inspection personnel must be used (for example due to travel restrictions). Such remote support scenarios can be expanded by engineers inside and outside of NDE. This can be taken to extreme by remote control of aerial drones or underwater robot-based inspection systems.

The recent Covid-19 pandemic, which led to global shutdown, forcing essential services to continue under stressful social distancing, demonstrated the value of remote NDE. In the next many months of low touch economy, NDE 4.0 enabling technologies is expected to significantly help protect the inspectors.

## *NDE for Industry 4.0*

**Quality Assurance in the Factory and the Infrastructure of the Future**

Smart manufacturing or the digital factory are technology-driven approaches that utilize Internet-connected machinery to monitor the production process. The goal is to identify opportunities for automating operations and use data analytics to improve manufacturing performance. And where is the data coming from? – #1 from NDE sensors, providing point-like information at short time intervals for in-situ inspections during manufacturing or operation, such as ultrasonic sensors (like vibration analysis or acoustic emission) and optical sensors (like IR, UV, visual), #2 from traditional NDE inspections, providing a view into the component at longer intervals. Only the combination of the data from both NDE sensors and inspections will provide the sufficient data input for predictive and prescriptive maintenance and structural health monitoring (SHM) of infrastructure such as buildings and transportation.

To enable this use case open standardized Industry 4.0 interfaces enabling semantic interoperability are required. "The NDE sector will not succeed in giving the industry new interfaces. It is more reasonable to use the Industry 4.0 interface developments and to participate in the design in order to shape them for the NDE requirements." [2,3]. Once the data is transferred it can be stored and used in digital twins, in digital threads, in data-base systems, or in clouds.

As manufacturing gets into mass customization, this would require NDE to adapt to the customized product, creating another use case for NDE 4.0.

**Quality Assurance of the Additively Manufactured Components.**

Components manufactured additively are usually difficult to inspect due to their complex internal structures or complex external shape. This is why the usability of most of the traditional NDE methods is very limited for those components. In most cases, out of the traditional methods, only computed tomography works.

This motivated several groups to start working on in situ NDE methods which monitor the signal during the additive manufacturing process. The most frequently used sensors are optical sensors monitoring and recording the internal and external dimensions using infrared, visual, or ultraviolet light. The heating and cooling processes, the melting and freezing processes, and the expansion and shrinking processes can be monitored. The feedback control can correct the process to ensure quality in real time.

NDE 4.0 can improve the 3D printing process. The reduced lot size of additively manufactured components is an additional challenge that can be addressed by NDE 4.0 much more conveniently.

**NDE of Drones and Critical Industrial Robots**

Very soon, society will be in the era where the expensive drones performing everyday functions will need inspection and maintenance programs. For most part, the society reacts to problems when they show up. It is not hard to imagine that one day in near future, a package delivery drone will have a catastrophic failure in someone's back yard and new regulations will emerge around inspection of aging drones and industrial cobots on similar lines. At the moment this technology and application is changing so fast that it is hitting obsolescence before aging, and so no maintenance plans are created.

**Continuous Improvement through Data Mining**

Once the data is transferred using standard interfaces it can be stored and used in digital twins, in digital threads, or in data-base systems; stored in the cloud. This NDE results/data becomes a valuable asset through statistical analysis and cross-correlations with other data sets. This asset can then be used for example in

- feedback loops for design improvements
- optimization of associated manufacturing processes
- trending to assure a constant or rising production performance
- probabilistic lifing methods to calculate the life of components more accurately [20]
- predictive and prescriptive maintenance to calculate the necessary maintenance inspections more accurately
- reliability engineering to enhance the reliability of components and products.

**Data Monetization**

New business models will emerge as data shows promise [2,3]. The structured data amenable to information extraction can become a commodity with a price tag, for data owners because it has value for product performance service life improvement. NDE 4.0 opens up the possibility of asset customized prescriptive maintenance, which can significantly improve the value derived from Data Analytics Maturity Model, originally proposed by Gartner in 2012. Analytics at various levels, require increasingly specialized skills, with possibility to command increased prices in the data market.

- Analytics Level 1: Descriptive – What happened?
- Analytics Level 2: Diagnostics – Why did it happen?
- Analytics Level 3: Predictive – What will happen?
- Analytics Level 4: Prescriptive – What should we do?
- Analytics Level 5: Cognitive – What don't we know?

## *NDE 4.0 Adoption Challenges*

In the previous sections, it was discussed how the digital technologies in NDE can help improve design, manufacturing, maintenance, and safety. So, what is holding it back. Well, the adoption still requires technology maturity, system robustness, a roadmap, a set of processes, and skills set, in addition to a financial capital, of course. The following will discuss these challenges, specific to NDE, before developing the strategy in line with design thinking.

### *Technical Challenges*

There are a few primary categories of technical challenges with the digital adoption, from revision of existing standards, to making the digital connectivity work, and demonstrating that system performs as promised.

**Existing Standards**

Standardization assures that the same mistake is not made twice but standards should not be used to prevent innovation. Some existing NDE standards were created years ago and they do not only cover the prevention of the "mistake" but also additional requirements describing a historic or outdated state of the art. This state of the art includes explicit requirements for analog equipment, like film for RT and analog UT or ET equipment, for handwritten and hand signed reports, for manually operated NDE equipment, and for visually performed data interpretation. This hinders innovation, prevents the implementation of NDE 4.0 and its use cases, and disables potential quality increase in inspection technology. Therefore, the existing standards need to be revised to accommodate new opportunities. Even more important today is to revise the standards development, acceptance, and governance process to enable adoption of rapidly changing technologies and business models.

**Digital Transformation of NDE**

The fourth revolution is data centric and lives from unhindered, secured data exchange between systems, programs, and assets from all kinds of organization within and outside of the NDE world. Industry 4.0 and NDE 4.0 require that burdens for digital communication are lowered, proprietary data formats and interfaces opened, and semantic interoperability implemented. The success of NDE 4.0 initiative requires open interfaces enabling data transparency and ensuring data security and sovereignty. It requires collaboration between different players on the market. This allows to combine technologies from different manufacturers and different industrial sectors. This allows manufacturers to focus on their core-knowledge which will finally result in better products which are more competitive. This allows both NDE to become part of the Industry 4.0 world and to use Industry 4.0 measures for NDE. This allows NDE and Industry to grow together as shown in Figure 1.

To enable the above-mentioned use cases open standardized Industry 4.0 interfaces are required. "The NDE sector will not succeed in giving the industry new interfaces. It is more reasonable to use the Industry 4.0 interface developments and to participate in the design in order to shape them for the NDE requirements." [2,3]. Basis for using the data (both in-situ or ex-situ) is semantic interoperability so that computer systems can understand the data. Once the data is transferred it can be stored and used in digital twins, in digital threads, in data-base systems, or in clouds.

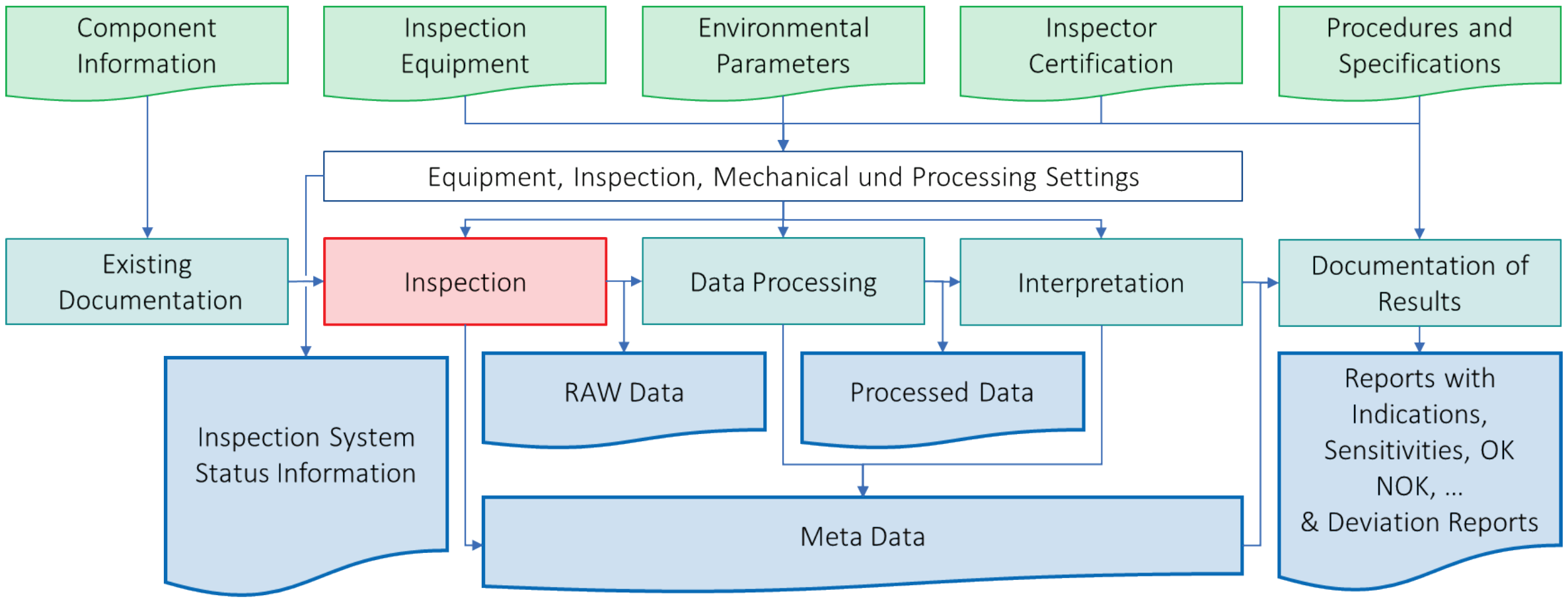

**Figure 5** Typical sequence of an (automated) inspection (can in principle be used for manual testing) [2,3].

[2,3] focusses on this topic, the digital transformation of NDE equipment, systems, and processes. Figure 5 shows the interfaces identified by [2,3] for manual, automated or automatic NDT / NDE inspections. For the use of NDE data for AI, AR, IIoT, Digital Twins, … the following output parameters can be used and semantic information models need to be established:

- raw data (for example HF UT Signals, raw ET or digital RT signals)
- processed data (for example C-Scans or Computed tomography)

- meta data (the settings of the inspection equipment including the settings of the data processing and interpretation)
- the reported values.

Highest value comes from the data that has already been interpreted. Therefore, the values reported in the NDT reports, the KPIs (Key-Performance Indicators) of an NDE inspection, needs to be made available first.

To integrate NDE into digital environments, for a completely digital inspection, also all the input parameters need to be digitized [2,3], like:

- Job & component information
- Inspection equipment
- Environmental parameters
- Inspector certification
- Standards, specifications, and procedures.

Job and component information and inspector certification can be provided by NDE Process 4.0 Management Tools as discussed above. Those tools need to implement standard interfaces to allow for such processes. Inspection equipment settings and in particular the digitization and digital interpretation of standards, specifications, and procedures to automatically program the NDE equipment and the mechatronics is a challenge which needs to be addressed by the scientific community.

**Standardization**

For those input and output parameters the semantic information models need to be standardized. This standardization can be accomplished, for example, by the creation of vendor-independent OPC UA companion specifications, or by the semantic information models provided by the DICONDE standards [2,3].

Moreover, standards need to be developed to assure a proper implementation of digital tools, like AI, automation, and the necessary basis, like data security and sovereignty. For example standards for NDE AI need to assure that such tools provide a reliability at least comparable to experienced human inspectors, that the human factors of AI training are highlighted, and that the training is performed appropriately and will not lead to constantly false data evaluation in later inspection scenarios.

**Validation**

NDE 4.0 application needs to demonstrate reliability, traditionally quantified by Probability of Detection (POD) or Intersection over Union (IoU). POD is hard from human factors perspective. So, on one side, the automation should make it very repeatable and dependable for known and deterministic applications. While on the other side, it is a challenge to quantify on newer application or deliver on the promise of adaptivity and agility of the learning system. POD protocols have to be developed and demonstrated for regulatory compliance and customer confidence. Machine intelligence adds a dimension that is not yet well understood from POD study perspective.

## *Non-Technical Challenges*

Data compiled and reviewed from series of focused discussions on the topic indicate that adoption of Industry 4.0 in general and NDE 4.0 in specific, face serious non-technical challenges in addition to the technical challenges around interface standardization and validation [15].

**Leadership**

There are several reasons holding decision makers from taking steps to create innovation-driven differentiation, starting with management incentives. Executives tend to be excessively risk averse, rightly so. The annual bonus programs promote data-driven incremental improvements. The promise of predictable near-term profit is a trap that so many managers still fall in, trading off larger intangible gains in the future for marginal visible gains in the ‘here and now’. NDE 4.0 is a long-shot game, requiring serious investment in technology and skills development.

Second, the management consultants who lead change initiatives and business school professors who publish bestsellers based on large amounts of data analytics generally provide valuable insight into successful companies; unfortunately, all in hindsight. The book Good to Great [24] became very popular in early 2000. Yet the growth model in 21st century has been very different, which is now captured in the book The Four [25]. Companies focus a lot of energy on “what” (metrics) and “how” (process), and more recently, “why” (purpose). Organizational consultant and writer Simon Sinek put these in the so-called Golden Circle [26]. However, adoption of NDE 4.0 requires a little different twist to the golden circle. Innovators start with “Why not?” and then go on to “How about?” and “What if?”

Third, NDE in certain industries is highly regulated. Regulations and innovation work in opposite directions. In general, the regulatory demands for compliance are not easy to meet, where innovations are revolutionary in nature with little to no precedence or data-based evidence to back up the value propositions.

Fourth, Industry 4.0 has brought about a business opportunity for a lot of technology solution developers, who may not fully appreciate the methodology, purpose, and intricacies of NDE 4.0. These solution providers are motivated to force fit what they sell to any customer they can have access to. Since, you may get approached by different providers with so many diverse options, you need an appreciation of what is a better fit and when. Lack of awareness around the technology, uncertainty associated with continuous change, and levels of funding required, make it very hard for management to decide where to invest.

Finally, the 4$^{th}$ industrial revolution requires a new style of leadership, that can tackle the issues surrounding people in *people-machine integration*. These leaders are digitally competent who can take responsibility of the people-side of this massive change, providing a clear direction and management in an open and transparent, employee-centric environment. Brian Bacon, Chairman and founder of the Oxford Leadership refers to this as Leadership 4.0 – “Leadership in the 4th Industrial Revolution will be defined by the ability to rapidly align & engage empowered, networked teams with clarity of purpose & fierce resolve to win” (https://www.oxfordleadership.com/leadership-4-0/)

Effective digital leaders in industry will be responsible for continuously changing interaction between technologies, machines and people, whilst nurturing ongoing knowledge-sharing, competency development, collaboration and innovation. They will need to mirror the technology of Industry 4.0 and IoT in that connectivity is at its core. Some call is connected leadership for that very reason.

**Skills and Competencies**

The inertia and friction associated with adoption of NDE 4.0 is the fear of job loss. Well, the revolution is coming, and like all previous revolutions, it will create more high paying jobs but for a different set of skills. Thus, a better way to look at is a skills and competency development problem rather than job loss problem.

The organizations will need a whole new skill set. Skills around Information and Communication Technologies (ICT not just IT), coworking with intelligent systems (desktop as well as industrial Cobots), and more importantly willingness to accept that what you know today will likely be obsolete before you can establish yourself as an expert. The need and speed for learning in the 4$^{th}$ revolution is an order of magnitude larger than the previous revolution. Employers and employees both need to take the learning and development as shared continual investment. While operators will need training on technology, the managers need to get on top of the processes, and leadership ought to explore new business models. The skill to rapidly learn and develop new skills will be the key. From within Industry 4.0, the AI reduces the need for operator technical training, augmented reality is enhancing training experience, and the cobots can be programmed in real-time through on the job execution.

**Management Processes**

Over the years, consistent focus on productivity improvement for near term profits has dampened the ability of companies to rapidly explore new space and create new products, or technologies or even exploit possible synergies across existing technologies. A very common workplace phrase is “we could do so much if we can get away from fire-fighting.” Exploration requires a bit of risk taking and investment along with a process that helps continuously assess and mitigate that risk. The term “Fail fast” or “Failure is a learning opportunity” requires a structured approach to work with. This is different than hit and trial, as discussed later.

## *NDE 4.0 Strategy*

With all the insight around technology, opportunity, and perceptions, the dots are now connected to define the purpose (why), roadmap (what), and process (how) to work through the next revolution in NDE [16].

### *NDE 4.0 Purpose*

Assuring safety is the number one motivation behind any asset inspection and maintenance. Everybody wants the system to function reliably, whether it is air, water, or a ground transportation vehicle; a material or energy manufacturing plant; a bridge or a building, an appliance or a piece of an equipment, and more. Everyone wants safety of all customers, users, stakeholders, operators, construction and maintenance crews, as well as the inspectors.

With 10 different uses cases at the intersection of NDE and Industry 4.0, the purpose of digitalization of NDE is unquestionable - provide an opportunity to advance all three - *quality(safety), speed, and cost*, as compared to the traditional perspective where you can only choose two out of the three [16]. That is why it is called revolution.

The improved and dependable POD provides *enhanced safety* and enables optimization of inspection programs reducing lifetime operating cost of an asset. The structured management of life time of digital thread, opens up additional *economic opportunity* to asset designers, manufactures, and operators.

This combination of *Stakeholder Safety* and *Economic Value* can be summarized under a single term **Safety 5.0**, shown in Figure 6, on lines similar to the definition of **Society 5.0** which brings economic value and social benefit through cyber-physical confluence of Industry 4.0.

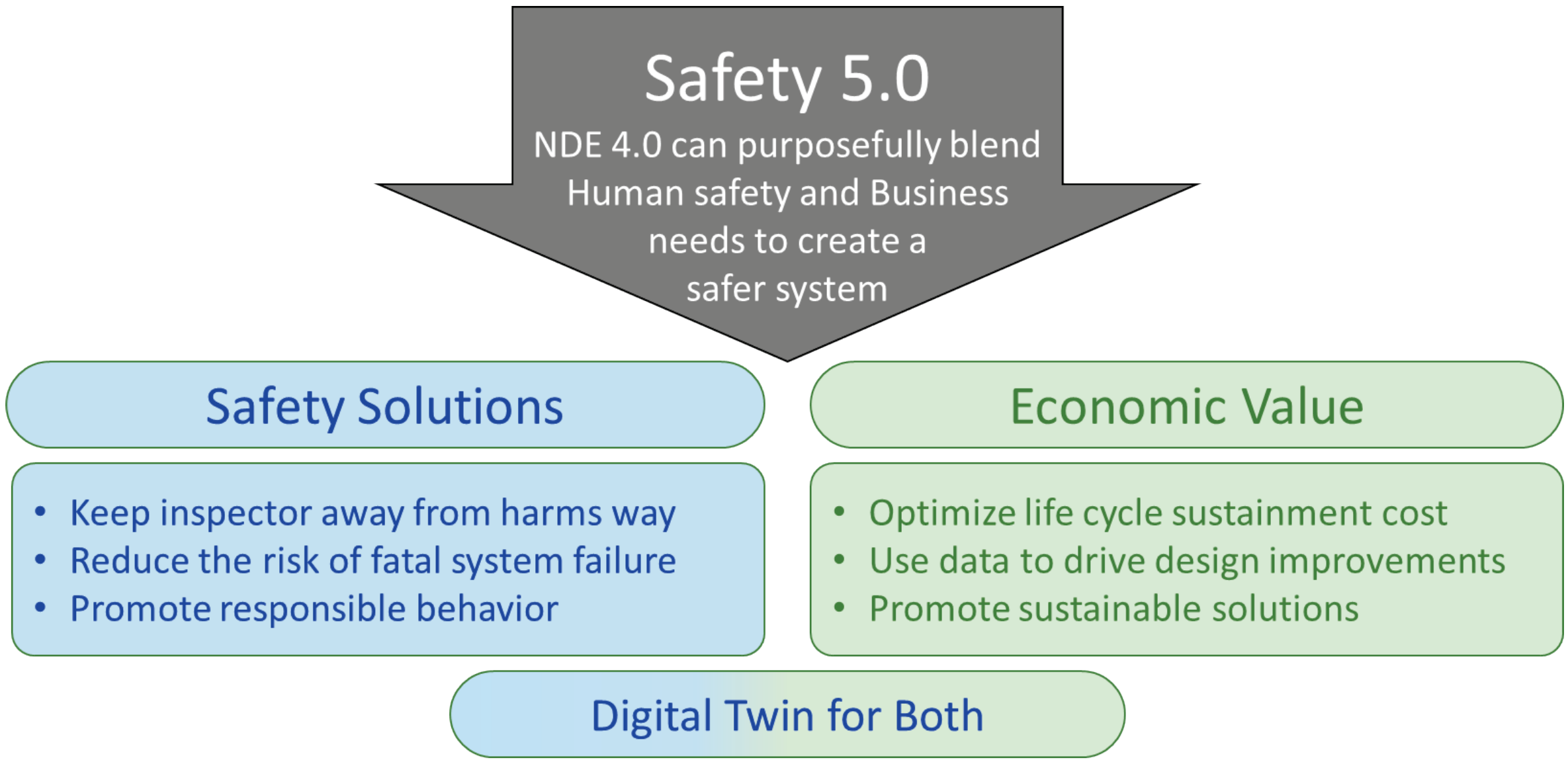

**Figure 6** Safety 5.0 as value proposition for NDE 4.0 [16] Other significant interests such as sustainability and sustainable development can be viewed as constraints or desirable outcomes.

It is also an opportunity to consider sustainability and sustainable development as a part of massive transformation under way. Humanity is beginning to realize importance of material life cycle and circular economy. NDE plays a major role in ability to repurpose, reuse ageing assets and infrastructure. NDE 4.0 makes it easier.

## *Adoption Roadmap*

As mentioned before, a revolution is not a discrete event that happens overnight. The digital technologies all emerge independently and then interdependently, and one day, a very different value proposition becomes apparent. The same is true for any organization. Various technologies discussed earlier need not be adapted all at the same time from technology value perspective. Nor is it affordable to do so from cash flow and talent development perspective.

Every business on a path to digital transformation needs a roadmap of what to adapt, when to adapt, and for what purpose. The roadmap needs to be aligned with market capture strategy on downstream side and talent/knowledge acquisition on the upstream side. It needs to account for evolution of the technological, economic, environmental, sociocultural, legal, and even political external environment The time span for roadmap is likely to be 7-10-year vision; 3-4 year strategic plan, and a 12-month tactical plan with allocated budgets. The attitude should be like a weather forecast. Hi fidelity and confidence for the near term, and a band of uncertainty for the far term; all of which needs to be actively managed with every passing quarter.  It is hard to define when the cross over to NDE 4.0 actually happens along the roadmap. It is just like, you start dating and living together, and one day you realize that you are in love; but can rarely point out the moment when transition happened.

Since NDE 4.0 is a revolutionary change, there is not enough consulting guidance on how to create a long-term workable roadmap. It needs an exploratory and learning mindset, and possibly a serious partnership across the eco-systems of NDE equipment manufacturers, digital experts, inspector trainers, and regulatory bodies; all facilitated by coaches who understand the purpose and the process.

## *Adoption Process*

Design Thinking is extremely useful in tackling problems that are ill-defined or unknown, by re-framing the problem in human-centric ways, creating many ideas in brainstorming sessions, and adopting a hands-on approach in prototyping and testing. Design Thinking also involves ongoing experimentation: sketching, prototyping, testing, and trying out concepts and ideas.

A standard and reliable process can reduce pain in all aspect of NDE 4.0 adoption cycle. It provides guidance on (a) how an organization can fulfil unmet inspection and safety needs, (b) enhance the competitiveness of organizations, NDE products, and inspection services, (c) lead to the easier acceptance of global inspection products, (d) reduce time to market for new inspection equipment, (e) increase the ability to make decisions, (f) take manageable risks, (g) face challenges and uncertainty associated with NDE 4.0, (h) evaluate the progress of the organization, (i) identify and share good practices in innovation management, and (j) share a globally accepted 'common language' for NDE 4.0.

### Simplified Approach for Ideation to Monetization or Value Realization

Innovation process for NDE 4.0 can be described as a series of connected steps, which leads to a new NDE product, service, system, or business model. Successful outcomes typically require an innovator to make assumptions at each step, validate or challenge them at the next step, go back if required, continuously build upon new learnings, and iteratively closeout on the purpose.

In its simplest form, the development project is about visualization of the end application, starting with a lot of ideas or opportunities; screening them down to a few projects; and then ethically executing them to a successful delivery or a new learning, sometimes; as shown in Figure 7. These steps are in line with Design Thinking philosophy.

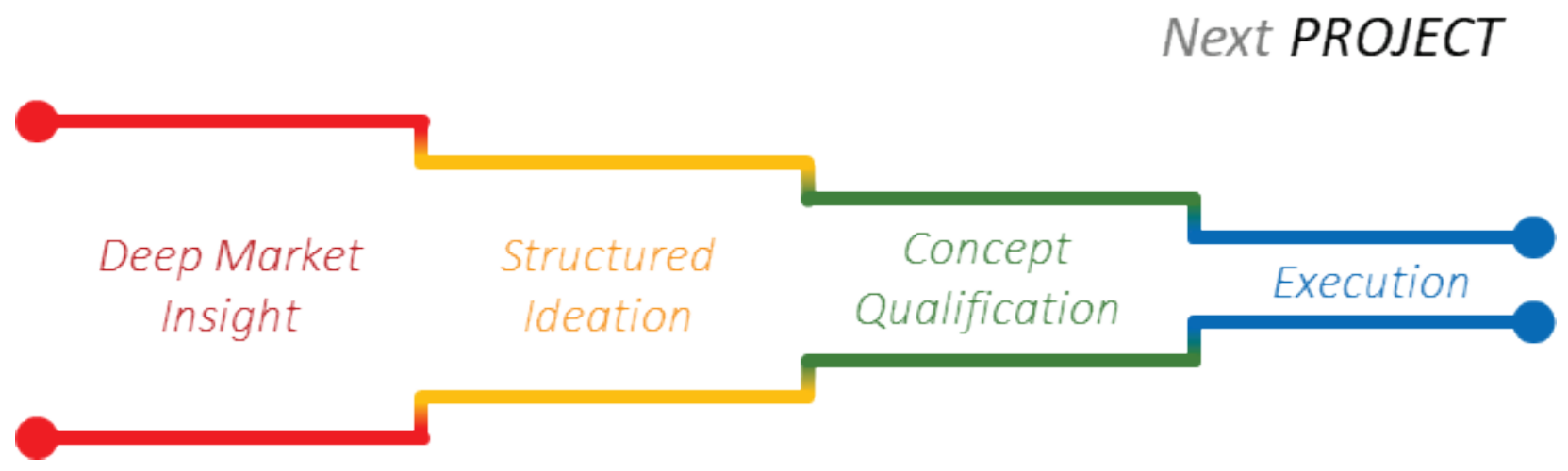

**Figure 7** Structured approach to innovation in context of NDE 4.0

### Market Insight

The first major step in NDE 4.0 adoption is to prioritize problems need to be addressed. The beginning of an innovation process is always very hazy. It is generally an accidental or deliberate matchmaking between an idea and the problem it can solve, sometimes difficult to separate, which came first. This is empathy and ideation to get to define stage.

When the objective is to take a new product or a service to the market, then there is a need for a good insight on what opportunities exist out there and on the competition. This marketplace insight is an absolute must to fulfill any desire or purpose to influence it. Design thinking has highlighted a number of opportunities and 'negatives' that NDE 4.0 needs to address.

NDE systems have two primary customers, inspectors, and owner/manufacturers/operators - all looking for zero job pain, at times conflicting with each other. The inspectors would like a system that is easy to learn, set up, carry, and operate, something that is physically comfortable: ergonomic, weather-proof, clean, with no messy consumables. The manufacturers, asset owners, and asset operators, ideally, want no inspections, but inspections are something they must live with for quality and safety assurance. So, they want minimum delays/downtime, maximum intervals between service inspections, and minimal cost of equipment, consumables, and so on. All of the stakeholders want reliable data and decisive information out of every inspection, desirable or not. The NDE 4.0 delivers to the requirements of both parties, making it a desirable activity, providing valuable data.

Complementary step is to see what already exists or is in works to solve the problem you have identified. Once again, there is generally a competition, or an incumbent solution resisting change to start with; and if you make it will invite new competitors to the market.

This step address first 2 stages of the design thinking process defined by d.school – Empathize and Define.

**Structured Ideation**

This is the most creative and elusive step of the innovation, and generally supports the myth that you can't teach innovation, or innovators are born. Ideation may not be science, but it is certainly not a magic. It is probably an art, and sustained practice gets you to a state of mind that is continuously generating great ideas. Experience has shown that the ideas are not a random occurrence, but rather triggered by some form of an intellectual stimuli. Which implies that 100s of ideas can be generated by using a stimulating environment or an exercise. Quite often, an ideation session can lead to new markets, in addition to serving a specific predefined objective. Invariably it must start with an objective. Typical Ideation objectives within NDE 4.0 can take the following form.

- What can AI or XR do for inspection systems?
- Which technology can solve the 'elusive problem'?
- How can the inspection activity be made safer for the inspector?
- How can NDE be accelerated by a factor of 10?
- What is needed to reduce the minimum certification levels for certain inspections?
- What lack of technology will make the business irrelevant in 5 years?
- How are regulatory requirements addressed, with limited evidence?

This step addresses the 3rd stage of the design thinking process defined by Stanford's d.school – Ideate.

**Purposeful Qualification**

After a couple of iterations between customer problem and solution ideas, these 3 questions must be addressed to assess the worth of the idea.

**(1) Value Proposition**: Does it add value to a customer/user/consumer? A good value proposition acts as a pain reliever and/or gain creator for a job to be done by a prospective customer. In this case it is the inspector and the asset owner/operator.
**(2) Purpose & Ethics Check**: Does it fit your self-defined purpose and self-imposed ethical standards, in additional to being legally compliant?
**(3) Concept Qualification**: Can you deliver it profitably, to sustain or grow your own business? Is it dependable, defendable, scalable, sustainable, and whatever else as a part of your purpose in the short or the long run? At times, you may have to iterate on a value proposition to qualify the idea.

There is no equivalent of this important step in the d.school process, which adds a significant value in the NDE 4.0 development activity, as it prepares the team to be successful at the next step of creating a prototype.

**Creative Execution**

A qualified concept now needs to be converted into reality. Most companies have some form of an R&D project management, or phase gate process to continuously reduce the execution and market risk. These include design reviews, lab testing, prototyping, first article acceptance, and regulatory approvals. Depending on how far out, your innovation is from existing knowledge and experience, you need to be prepared to iterate on value proposition and execution options. The ability to learn and adapt is still the key to successful innovation.

This is broader and more comprehensive than steps 4 & 5 in the d.school process – Prototype & Test.

**Value Chain is a Multilayer Filter**

This activity from ideation to monetization is like a multilayer filter. At each step, you remove the options that may not work. Typically, out of 100s of ideas only a handful will qualify for execution. At this stage, you could either create a portfolio of projects, or use a criterion for prioritization.

Marketing folks might see this as a funnel. Which is actually a bad metaphor because in a funnel everything that gets in from the top gets out at the bottom. In a well-designed filter, only the desired material comes out. In case of innovation value chain, the filter design is innovation management processes and the human mindset.

The simplified '*ideation to monetization*' innovation value chain, described in here can be used on routine basis to address the growth needs of an organization or as a core execution engine in any of the NDE 4.0 applications.

The steps will be iterative, more often than desired. It takes a bit of a practice to get good at it, just like playing piano.

### *ISO 56002:2019 Innovation Management System – Guidance*

A more generic guidance has just emerged from International Organization of Standards. The ISO 56000 series of documents on innovation management guidance can be used for conception, development, validation, and pursuit of purposeful NDE 4.0 applications. *ISO 56002* [27] document provides guidance for the establishment, implementation, maintenance, and continuous improvement of an innovation management system for use in all established organizations. It is applicable to NDT equipment manufacturers, service providers, training schools, and asset owners responsible for infrastructure safety. All the guidance within this document is generic and intended to be applicable to all types of organizations, regardless of type, sector, or size; all types of innovations, e.g. product, service, process, model, and method, ranging from incremental to radical; and all types of approaches, e.g. internal and open innovation, user-, market-, technology-, and design-driven innovation activities.

*Relevance*: The ISO standard is an overarching document that integrates all of the remaining ISO 56003-08 documents on innovation, which refer and eventually feed into for successful execution. It does not describe detailed activities within the organization, but rather provides guidance at a general level. It does not prescribe any requirements or specific tools or methods for innovation activities. This intent makes the application as broad as possible, including NDE 4.0, which also subjects the user to differences in understanding and interpretation. It is directly relevant, with some level of understanding of the overlap across fundamentals of NDE and basics of Innovation.

### *Skills Development*

As discussed before, most people view Industry 4.0 as a job killer. It is not a jobs question. It is a skills question. Jobs will be there. People will have to upskill themselves.

From an organization perspective, this includes Leadership 4.0 skills, Industry 4.0 skills, as well as an open mindset which strikes a balance across exploration and exploitation. Individual talent development plans need to support the technology awareness and adoption roadmap. There are a number of open online courses and university certification programs that can be leveraged. Talent development needs to be viewed as a shared investment by employee and employer to make it work for both parties.

### *Adoption is a Journey*

There may not be a single discreet event that defines the cross over to NDE 4.0. It is a journey, and most are already on it. Any new awareness through this paper is the step in that journey, which is different for different organizations and individuals.

From NDE technology/equipment/systems perspective, various stages correspond with, respectively

1. seeing what is happening (data),
2. knowing why it's happening (analytics, knowledge),
3. predicting what will happen (based upon the patterns and capabilities developed before and AI) to the ultimate step Industry 4.0 strives for:
4. an autonomous reaction by autonomous machines within the self-optimizing NDE 4.0 systems.

An example of a major milestone is the first instance of digital twin based full loop cyber-physical integration of an inspection. From there it can continue to spread internally, externally and number of connections. Perhaps, you could claim successful adoption when more than 50% of the decisions are made by the machine or 50% of the inputs and outputs are digitalized.

From a business perspective, a good milestone would be when the organization begins to see both the economic value and stakeholder safety. Remember this combination won't last very long. The transformation journey ends when one needs to choose between cost, speed, and quality.

### *Adoption needs Community Engagement*

NDE 4.0 development needs an eco-system approach. It is so intricate in terms of development and adoption, that no single organization can claim ownership or leadership, worth claiming based on impact. Universities, corporate R&Ds, equipment manufacturers, asset manufacturers, asset owners, inspectors, inspection service providers, asset consumers, trainers, regulators, and professional bodies have to come together to identify their roles for a purpose. The African proverb fits here – "*it takes a village to raise a child*" which means that an entire community of people must interact with children for those children to experience and grow in a safe and healthy environment.

The NDE community is converging on a need for cooperation. With the formation of the DGZfP (German Society for NDE) committee "ZfP 4.0" in 2017, of the ICNDT (International Committee for NDT) Specialist International Group "NDE 4.0" in 2018, of the ASNT (American Society for NDT) committee "NDE 4.0" in 2019, and of the BINDT (British Institute for Non-Destructive Testing), AEND (Spanish Association of Non-Destructive Testing), APFNDT (Asia Pacific Federation for Non Destructive Testing) committees in 2020 the NDE industry started activities regarding NDE 4.0. An NDE 4.0 ambassadors group is meeting in regular intervals for international exchange, multiple special issues on NDE 4.0 have been started by Materials Evaluation, RNDE, and the Journal for Nondestructive Evaluation, and a YouTube channel [18] as well as a LinkedIn group have been started to bring awareness on this topic. The first DGZfP seminar on "ZfP 4.0" will be October 8$^{th}$, 2020 and the first international conference on NDE 4.0 will be April 2022, with a virtual precursor in spring 2021.

## *Discussion and Outlook*

The global NDE / NDT community is at such an early stage of conversation around NDE 4.0 that it is not prudent to make any conclusions or closing remarks. Therefore, a discussion and outlook are provided.

Anything purposeful is worth pursuing.

This paper has addressed multiple aspects to the question "What is NDE 4.0 and what is its value proposition?" Like with Industry 4.0 the answer is not easy or straightforward at first glance. The use of the emerging digital technologies for NDE and the new possibilities for inspection control (Industry 4.0 for NDE) are major use cases. The other major use cases identified are "NDE for Industry 4.0". All of the uses case identified so far show that Industry and NDE are growing together with the fourth revolution.

All the use cases described will eventually lead to an improved awareness about NDE. This will help to complete NDE sector and it will help to increase the value people see in NDE.

There is still a long way to go. When the barriers to digital communication are lowered, proprietary data formats and interfaces are replaced, and semantic interoperability is natural; then it will be possible to combine the emerging technologies into a new cyber-physical inspection equipment. It will be possible to connect equipment from different manufacturers, and analyze the big data for safety and quality. It will enable manufacturers to focus on their core-knowledge resulting in rapidly improving products and superior services.

Given the challenges and opportunities, the NDE 4.0 needs collaboration on an international scale, without burdens or old structures. Ideas like NDE-manufacturer based clouds are the use of emerging technologies for maintaining the old structures. This eventually may not work. Opening up, collaboration and the willingness to innovate are key to NDE 4.0 and will decide on the future of individual companies and of the NDE sector in total.

If taken on thoughtfully, the NDE 4.0 will lead to a completely new way of sustaining product quality and safety, a new way of doing business, a new market for data – an ecosystem with huge potential for purposeful NDE.

## *Acknowledgement*

Many thanks to Mark Pompe, Norbert Meyendorf, and Anish Poudel of ASNT for supporting the NDE 4.0 activities and creation of the ASNT committee, to the anonymous participants of the survey on Facebook and LinkedIn, and to the participants of the NDE 4.0 class at the ASNT Annual Conference 2019.